\documentclass[a4paper,showkeys,floatfix,aps,pre,longbibliography,superscriptaddress,preprint]{revtex4-1}

\usepackage[dvipsnames]{xcolor}
\definecolor{linkcolor}{rgb}{0.3,0.3,1.0} 

\usepackage{amsmath,amssymb}
\usepackage{graphics,graphicx}
\usepackage{dcolumn,bm}
\usepackage{color}
\begin{document}

\title{Social dynamics through kinetic exchange: The BChS model}
\author{Soumyajyoti Biswas\\ Department of Physics, SRM University - AP, Andhra Pradesh 522240, India \\  Arnab Chatterjee \\ TCS Research, New Delhi, India \\  Parongama Sen \\ Department of Physics, University of Calcutta, Kolkata 700009, India\\  Sudip Mukherjee \\ Barasat Government College, Barasat, Kolkata 700124,  India \\  Bikas K. Chakrabarti \\ Saha Institute of Nuclear Physics,  Kolkata 700064, India \\ Indian Statistical Institute, Kolkata 700108, India  }


\begin{abstract}
This review presents an overview of the current research in kinetic exchange models for  opinion formation in a society. 
The review begins with a brief introduction to previous models and subsequently provides an in-depth discussion of the progress achieved in the Biswas-Chatterjee-Sen model proposed in 2012, also known as the BChS model in some later research publications.
The unique feature of the model is its inclusion of negative interaction between agents.
The review covers various topics, including phase transitions between different opinion states, critical behavior dependent on various parameters, and applications in realistic scenarios such as the United States presidential election and Brexit.
\end{abstract}


\maketitle
\section{Introduction}
It has been a popular notion for a while to perceive human society as a complex network system~\cite{oup,RevModPhys.81.591,galam_book}. The actions of individual human beings, as they interact with one another through social or economic network links, lead to a social non-equilibrium steady state with macroscopic characteristics such as wealth distribution, opinionated consensus, etc. This is similar to how a many-body condensed matter system arrives at an equilibrium macro-state through interactions. However, there are criticisms and challenges to this idea. One significant issue is that the individual constituents or ``social atoms" are themselves quite intricate and may not adhere to the straightforward, well-defined laws of interaction assumed in models of physical atoms in ideal gas or material systems.

However, in the appropriate context,  the interactions between the individuals and the resulting changes in the values of the social variables (e.g., money, opinion, etc.) can be sufficiently restricted so that the complexities arising out of the presence of self-deciding individuals rather than well-defined gas molecules, are minimized. For instance, the nuance involved in one's political opinion gets drastically reduced when at the polling stations they have to choose between predominantly two opinions (say, in Brexit).  It is, therefore, a valid context to consider binary opinion values for the individual agents, or even a generalization towards having a continuous range of values between two extreme ends. 

\begin{figure}
\includegraphics[width=0.5\linewidth]{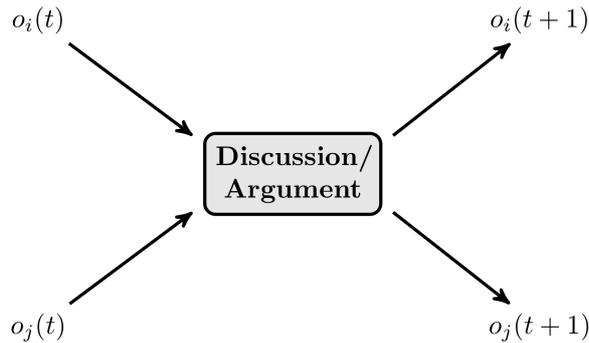}
\caption{A schematic representation of the kinetic exchange opinion model. Two agents, $i$-th and $j$-th, come to a discussion/argument at the time $t$ with the respective opinion values $o_i(t)$ and $o_j(t)$. After the discussion, they modify their opinion values to $o_i(t+1)$ and $o_j(t+1)$. The modification process is a linear relation with generic form given in Eq. (\ref{gen_eq}), however a non-linearity enters through enforcing the bounds $|o_k(t)|\le 1$ on the opinion values (real numbers).}
\label{schemaic}
\end{figure}

The complexity (or assumptions, or interests) then translates into formulating the interactions between the agents. This is where a class of models were formulated (see e.g., Ref.~\cite{doi:10.1142/S0219525900000078}) which consider the interaction between the agents as an exchange of opinions between the individuals~\cite{PhysRevE.82.056112}.  Mathematically, if the opinion value of the $i$-th agent at an instant of time $t$ is denoted by $o_i(t)$, then it could evolve following
\begin{equation}
    o_i(t+1)=f(o_i(t),o_j(t)),
    \label{gen_eq}
\end{equation}
where an interaction/exchange has happened between the $i$-th and the $j$-th agents and the function $f$, which represents this interaction/exchange process, is a linear function of its arguments. There is, however, a non-linearity in this process that comes from the fact that the opinion values of all agents are bounded ($|o_i(t)|\le 1$)
at the extreme values ($\pm 1$) of opinions.
This simple linear form is inspired by the similar genre of models of wealth exchanges in a closed economy~\cite{cc,ccm}, which was in-turn inspired by the kinetic theory of ideal gases (see Fig. \ref{schemaic} for a schematic diagram). An `exchange' in this context is mostly the formal similarity with wealth exchange and also is in the spirit of exchanging ideas or information between the selected pair that could then lead to shifts in their opinions.  

Like the kinetic theory of ideal gases, the exact form of interactions is hard to track down but as is known in the kinetic theory that the departures so introduced, are averaged out in the statistical sense.  Unlike the kinetic theory, of course, the interactions are not just  exchanges of energy (wealth or opinions), but instead are accompanied by a saving propensity that keeps a fraction of the exchanging quantity for themselves and "trade" with the remaining part. While a conservation of wealth is still obeyed in wealth exchanges, there is no such conservation for opinion exchanges. 
Instead, following an exchange of opinions, the two participating individuals either come closer together in their views or drift apart, depending on the nature of the ``exchange" they had between them. 
As a result, the collective opinion of the society can either shift towards an emergent consensus or can get fragmented. 
Interesting questions arise regarding the conditions that facilitate emergent consensus, such as the effects of topologies (i.e., the way agents are connected and interact), the impact of non-conformist individuals on global consensus, and the time required to reach a consensus, the proximity to breakdown of consensus, and so on.
Another line of investigation is on the characterization of the transition between consensus and fragmented state in the form of a critical phenomenon. The exponent values and the corresponding universality class are of interest.

In this review, we will first discuss the formulation of the kinetic exchange models for opinion dynamics and what do the parameters of the models mean for various different features of opinion exchanges between individuals (Sec.IIA). We will then move onto the phase transition behavior seen in such models, the nature of the phase transitions, and the different variations of the models where the individuals include non-conformists. Then we will discuss the effect of topology on such phase transition behavior (Sec. IIB). Finally, we will discuss the various different situations where such models could be applied -- the case of the US presidential election (Sec.IIIA), the case of Brexit (Sec.IIIB) and some models of tax evasions (Sec.IIIC), for example. The corresponding comparisons with the real data were discussed wherever possible, and then we summarize the results discussed and provide the outlook.

\section{Kinetic theory of social exchanges}
As noted in the Eq. (\ref{gen_eq}), the evolution of the opinion values $o_i(t)$ follows a linear exchange with the opinion value of another agent $o_j(t)$, but the resulting process could be non-linear, in order to incorporate the bounds at the extreme values $\pm 1$. 
 The values of $o_i(t)$ can be either continuous within this range, or discrete ($\pm 1,0$) that includes a neutral opinion explicitly.  

 Obviously, the interaction/exchange in the above-mentioned scenario is a complex process, but we argue that the crux of the resulting reshaping of the opinion values could be captured by relatively simpler rules in a statistical sense, i.e., the departures from such simple rules cancel out on average. Of course, this is a simplifying assumption. In this case, we assume that a particular individual retains a part of their original opinion state (unlike in the voter and related models) and is influenced only partly by the opinion state of the other individual, hence the exchange. Particularly, 
\begin{equation}
    o_i(t+1)=\lambda o_i(t)+\lambda\epsilon(t) o_j(t),
 \label{lccc_dyn_eq}
\end{equation}
where the ``exchange" is considered between the $i$-th and $j$-th agents. We will refer to this as the LCCC model, which was introduced in Ref.~\cite{PhysRevE.82.056112}.  During the interaction (written here from the view of the $i$-th agent, and a similar equation could be written for the $j$-th agent as well), the agent retains $\lambda$ fraction of their original opinion (at time $t$) and gets influenced by the $j$-th agent, such that $\lambda\epsilon(t)$ fraction of the $j$-th agent's opinion is added to it. Here $\lambda$ is a constant across all agents and $\epsilon(t)$ is a random variable drawn at each time independently for each interaction from a uniform distribution in $(0,1)$. There is no restriction on the choice of $i$-th and $j$-th agent. However, a bounded confidence variant was studied in Ref.~\cite{PhysRevE.86.016115}, where the agents interact only when the difference between their opinion values remain within a limit. 
\begin{figure}
\includegraphics[width=0.6\linewidth]{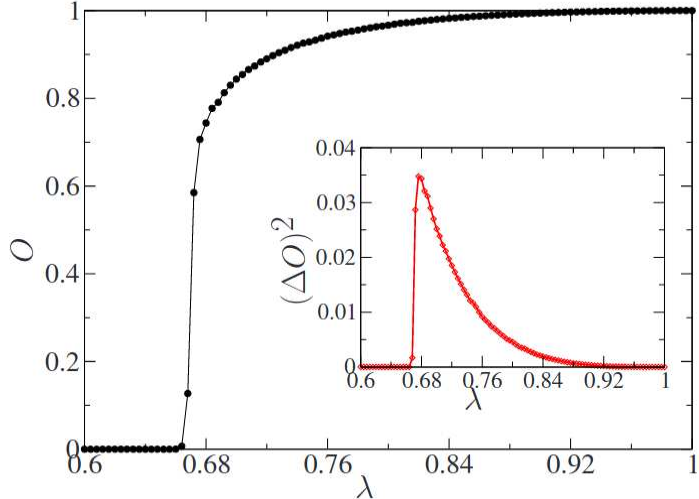}
\caption{For the LCCC model, Monte Carlo simulation results for the order parameter $O$ is plotted as a function of the conviction parameter $\lambda$.  The inset shows the same for the
fluctuation in $O$. The symmetry breaking transition is seen at $\lambda_c$, which is approximately $2/3$. Taken from Ref.~\cite{PhysRevE.82.056112}.}
\label{op_schemaic}
\end{figure}
Note that the interaction process here is such that it is non-negative, meaning that if the two agents had belonged to the same side of the opinion spectrum (both positive or both negative), then after the interaction they would remain on the same side. This is easily seen if the above equation is rewritten as
\begin{equation}
    \frac{o_i(t+1)}{o_j(t)}=\lambda\frac{o_i(t)}{o_j(t)}+\lambda\epsilon (t).
    \label{lccc_dyn}
\end{equation}
Given that in the above equation the last term is positive definite, if the other term on the right-hand side is positive then the left-hand side must also be positive. Therefore, there is a spontaneous symmetry breaking transition for sufficiently large values of $\lambda$, where all opinion values are of the same sign (see Fig. \ref{op_schemaic}). For low values of $\lambda$, all opinion values eventually become zero. The nature of the symmetry breaking transition in this model has been investigated widely. Within the framework of the critical phenomena, simulation results indicate that it does not belong to the Ising universality class (even though it breaks a $Z_2$ symmetry), or that of mean-field active-absorbing transition (although the system reaches an absorbing state below a critical value of $\lambda$). The order parameter is defined as the average of the overall opinion values
\begin{equation}
    O(t)=\frac{1}{N}\left|\sum\limits_io_i(t)\right|.
\end{equation}
In the steady state (long time limit), near the critical point ($\lambda=\lambda_c \approx 2/3$), it fits a variation of the form
\begin{equation}
    O(t\to\infty)\sim (\lambda-\lambda_c)^{\beta},
    \label{op_lccc}
\end{equation}
with $\beta\approx0.1$~\cite{PhysRevE.82.056112}. This value of the order parameter exponent is significantly less than that of the mean-field Ising model ($\beta =1/2$) or the mean-field active-absorbing transitions ($\beta = 1$). Additionally, there is very little system size dependence in this version of the model.

As mentioned earlier, an equation similar to Eq. (\ref{lccc_dyn_eq}) could be written for the $j$-th agent as well. As far as the numerical simulations of these models are concerned, the steady-state properties do not depend on the types (asynchronous or synchronous) of updates.

A mean field calculation was proposed~\cite{Biswas_2011} for the fixed point $o^*$ given by
\begin{equation}
o^*[1-\lambda (1+ \langle \epsilon \rangle)] = 0,
\end{equation}
from where it follows that the critical point $\lambda_c = 1/(1+ \langle \epsilon \rangle)$ where $\langle \ldots \rangle$ refers to average.
For uniform random distribution of $\epsilon$, $\langle \epsilon \rangle =1/2$ and hence, $\lambda_c = 2/3$. Here, it is important to note
that this mean-field treatment does not incorporate the cut-offs at $\pm 1$. It was also noted that the underlying topology (1d, 2d or infinite range) has barely any effect on the critical point.

A variant of this model was later proposed~\cite{PhysRevE.83.016108} where the conviction parameters of the agents and the parameter representing the influence of the others were taken as different. The exchange equation then reads
\begin{equation}
    o_i(t+1)=\lambda o_i(t)+\mu \epsilon(t) o_j(t).
\end{equation}
Here, the behavior of the model is non-universal along the $\lambda-\mu$ plane, with the original model being recovered at the $\lambda=\mu$ point. 

Various attempts have been made to seek analytical solutions for this category of models, all of which demonstrate very little fluctuations with respect to system size and undergo a spontaneous active-absorbing type of symmetry breaking transition. However, they maintain a distinct set of critical exponent values that are far from the anticipated mean-field class of active-absorbing transitions in models of this nature~\cite{Biswas_2011}.

One such attempt~\cite{10.1007/978-3-319-00023-7_7} was to write Eq. (\ref{lccc_dyn}) in the form of a mean field like dynamical evolution of the form
\begin{equation}
    O(t+1)=\lambda(1+\epsilon(t))O(t),
    \label{map}
\end{equation}
where $O(t)$ represents the mean field average opinion value. 
One can study the stochastic map  in Eq.~(\ref{map}) by describing it in terms of random walks.
Writing $X(t)=\log({O(t)})$ (for all subsequent discussions we always take $O(t)$ to be positive), Eq.~(\ref{map}) can be written as
\begin{equation}
X(t+1)=X(t)+\eta,
\label{rw}
\end{equation}
where, $\eta(t)=\log[\lambda(1+\epsilon)]$. As is clear from the above equation, it actually describes a random walk 
with a reflecting boundary at $X=0$ to take the upper cut-off of $O(t)$ into account. Depending upon the value of
$\lambda$, the walk can be biased to either way and is unbiased just at the critical point. As one can   average independently 
over these additive terms in Eq.~(\ref{rw}), this gives an easy way
to estimate the critical point~\cite{PhysRevE.82.056112}. An unbiased random walk would imply $\langle \eta\rangle=0$ i.e.,
\begin{equation}
\int\limits_0^1\log[\lambda_c(1+\epsilon)] d\epsilon=0
\end{equation} 
giving $\lambda_c=e/4 \approx 0.68$, where a uniform distribution of $\epsilon$ in the range (0,1) had been considered.
The tricky averaging here over
the log function may be performed using the transformation $x =
\log(1 + \epsilon)$, giving $\int_0^1 log (1 + \epsilon) d\epsilon
= \int_0^{log2} x e^x dx = 1 - 2log2$. 
The steady state value of $O(t)$, which is the equivalent of the order parameter in the multi-agent model, turns out to be of the form
\begin{equation}
{O(t\to\infty)}= \exp[-{k}|\log\lambda|^{3/2} (\lambda-\lambda_c)^{-1/2}].
\label{averageO}
\end{equation} 
This is not a power-law variation of the form taken in Eq. (\ref{op_lccc}).

A further modification was proposed~\cite{PhysRevE.84.056106}, where the values of the conviction parameter $\lambda$ were made stochastic, in the sense that $\lambda=1$ with probability $p$ and $\lambda=0$ otherwise. This modification makes the model analytically tractable, because the opinion values now become discrete ($\pm 1, 0$) if one starts from discrete initial conditions. The non-negativity of the interactions discussed above results in a polarization (either positive or negative opinions survived) in the system that could be shown analytically.   It was then shown that the order parameter, in the steady state limit ($t\to\infty$) behaved as
\begin{equation}
    O(t\to\infty)=\frac{3(p-2/3)}{p},
\end{equation}
implying an order parameter exponent $\beta=1$, consistent with mean field active-absorbing transition. Also, the finite size scaling and the related exponents were observed from an off-critical scaling of the form
\begin{equation}
    O(t)=t^{-\delta}\mathcal{F}\left(\Delta t^{1/\nu_{||}},\frac{t^{d/z}}{N}\right),
\end{equation}
where $\nu_{||}$ is the time correlation length exponent, $z$ is the dynamical exponent, $d$ is the space dimensionality, $N$ is the system size and $\Delta=p-p_c$ the critical interval. The scaling relation $\delta=\beta/\nu_{||}$ is maintained here, where all the  values are close to unity. Interestingly, if $d=4$ is assumed (as the upper critical dimension), then $z=2$ is obtained, as is expected for mean field. 

It is interesting to note that if the two-agent exchange condition were to be relaxed, the transition behavior changed significantly. For example, a three-agent interaction was considered in the following way: three agents were chosen at random, then the first agent interacted with the other two only if the opinion values of the other two agents matched. Keeping other parts of the dynamics same, this results in a discontinuous transition, where the order parameter behaves as $O(t\to\infty)=\frac{1}{2}+\frac{3\sqrt{p-8/9}}{2\sqrt{p}}$. For a mixture of the two-agent and three-agent interactions, the transition eventually becomes continuous, passing through a tri-critical point at a critical value of the mixture. 

However, it was also noted that the nature of the `disordered' phase in all the above-mentioned versions is peculiar in the context of opinion formation, so as to have all opinion values at zero, meaning a neutral phase. Usually, in a society with competing opinions, the disordered phase is a fragmented one with almost equal sizes of population on either side of the issue (and possibly with few neutral agents). The reason for this `absorbing' phase in the disordered state of these models is the non-negative nature of the interactions discussed above. Indeed, even for the ordered phase, opinion values of one sign survives~\cite{PhysRevE.84.056106}.  Therefore, a negative interaction, in the form of a noise, was subsequently introduced in the model.

However, before going to the discussions on negative interactions,
let us first mention the universality of the LCCC model under the influence of an external noise~\cite{MUKHERJEE2021125692} (see also Ref.~\cite{Freitas_2020}, for effect of external field). The external noise can be incorporated in the dynamics as follows:
\begin{equation}
    o_i(t+1)=\lambda o_i(t)+\lambda\epsilon(t) o_j(t)+\eta_i
\end{equation}
where $\eta_i$ is a random uncorrelated noise that can be either $+1$ or $-1$. This would destroy the absorbing nature of the disordered phase, yielding a co-existence of the opinion values of both signs in the disordered phase. The critical exponents in that case turns out to be close to those for the Ising model.

\subsection{The BChS model}
In Ref.~\cite{BISWAS20123257}, a version of the kinetic opinion exchange model with negative interaction  was introduced, where the transition was governed by a
tunable noise. Following the subsequent
studies and following a naming of the
model (see,  e.g., Refs.~\cite{RAQUEL2022127825,e25020183,ALVES2021125267,LIMA2021125834}), the model is
called here by the name BChS model.  The
evolution in the model follows binary
exchange between the randomly selected $i$-th and $j$-th agents, with the evolution rule
\begin{equation}
    o_i(t+1)=o_i(t)+\mu_{ij}(t)o_j(t),
    \label{bcs_dyn}
\end{equation}
 with no sum over $j$ implied. If an extreme end ($\pm 1$) is reached, then the opinion values are kept fixed at the extreme value. Here, the parameter $\mu_{ij}$ signifies the interaction or an `opinion relationship' or `alignment index' between agents $i$ and $j$, and thus the opinion after interaction depends on the nature of this relationship as well as the instantaneous values of the opinions of the pair.
 $\mu_{ij}$ is generally taken as independent of $i,j$, and takes the value $-1$ with probability $p$ and $+1$ with probability $1-p$. Clearly, this is a noise parameter that allows a negative interaction, i.e., two agents could be on one side of an issue (having the same sign of the opinion values) but could end up on different sides of the issue (having different signs of the opinion values) after the exchange. The parameter $p$ simply describes the probability of opposing relation that a pair of agents have at that particular exchange. Since this is an annealed variable in general, but for the mean field case the nature of this variable (quenched/annealed) is irrelevant. 
\begin{figure}[t]
\includegraphics[width=7.5cm]{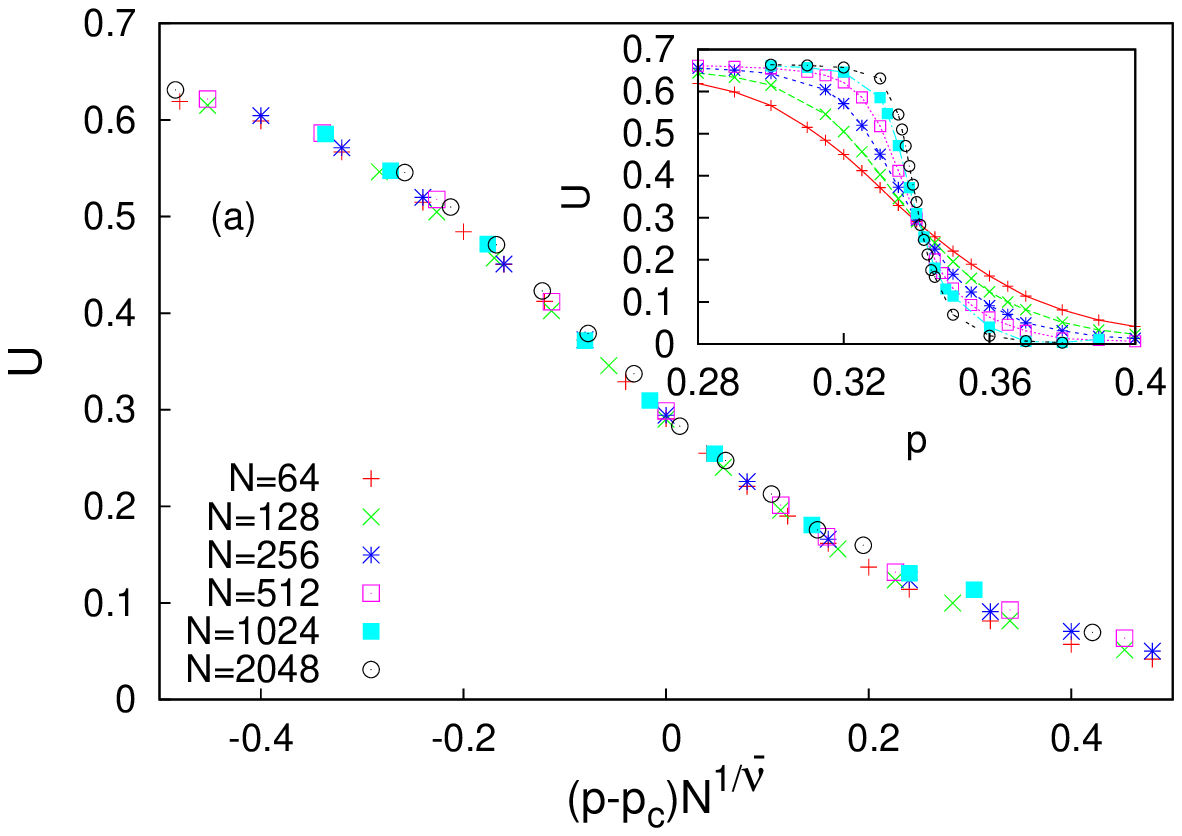}
\includegraphics[width=7.5cm]{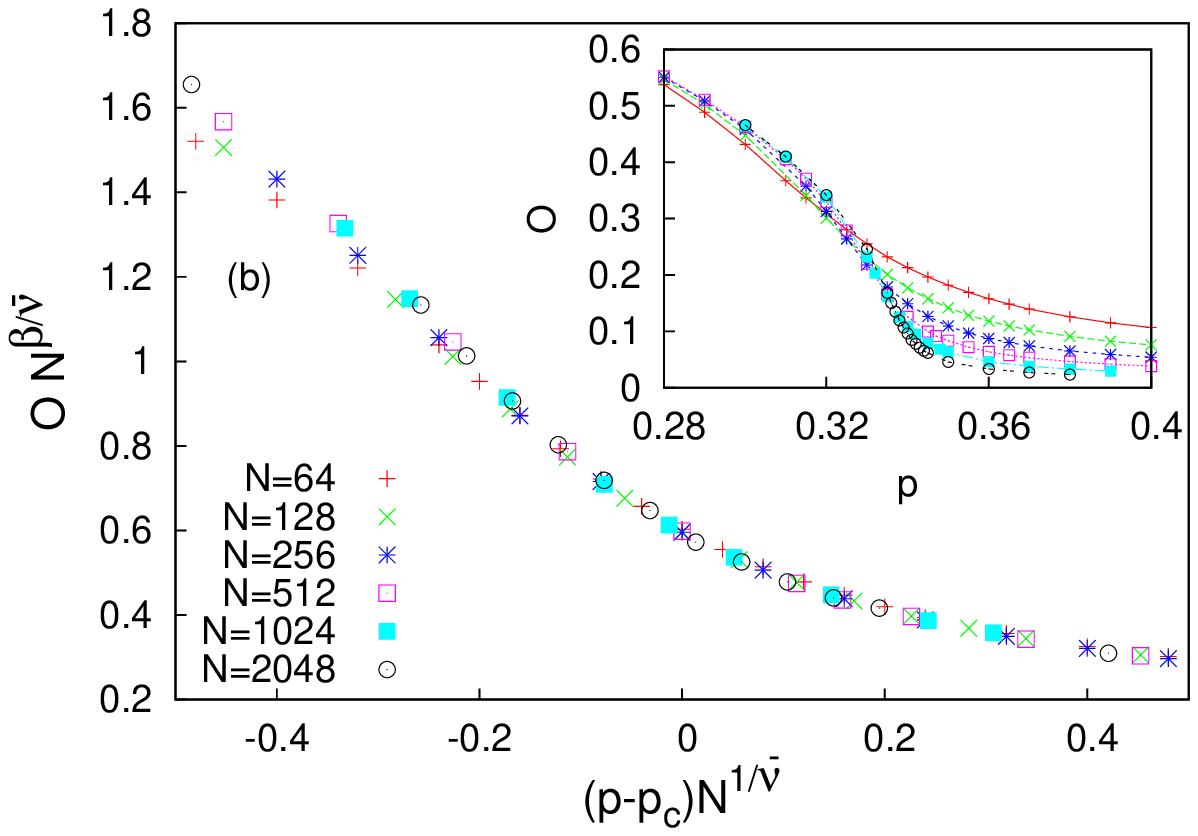}
\includegraphics[width=7.5cm]{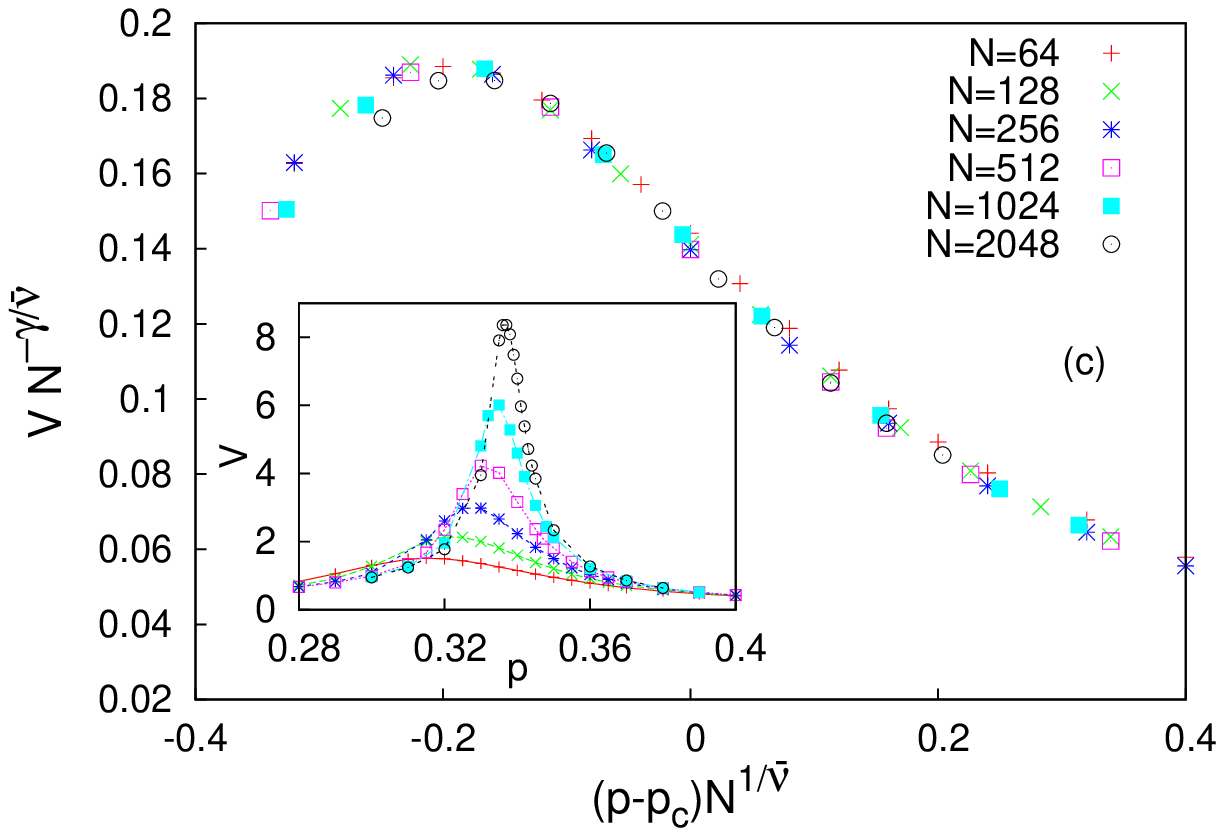}
\caption{Numerical simulation results for the BChS model with continuous, annealed $\mu_{ij}$ model, showing
(a) finite size scaling of the Binder cumulant $U$ for various system sizes $N$;
 the critical point is $p_c = 0.3404\pm0.0002$,  with the best data collapse for $\overline{\nu}=2.00\pm 0.01$.
Inset: $U$ as $p$ varies for different system sizes; 
(b) finite size scaling of the order parameter $O$ for various system sizes $N$;
 best data collapse is for 
$\beta=0.50 \pm 0.01$.
Inset: $O$ as $p$  varies for different system sizes;
(c) finite size scaling of $V$  for various system sizes $N$; the best
 data collapse is for 
$\gamma=1.00 \pm 0.05$.
Inset: $V$ as $p$  varies for different system sizes. 
The number of averages are 
 3000  for $N=256$, 1800 for $N=512$, 1000 for $N=1024$ and $400$ for $N=2048$. Taken from Ref.~\cite{BISWAS20123257}.}
\label{fig:contannU}
\end{figure}
 In the mean field limit (any agent can interact with any other agent), the dynamics are analytically tractable, particularly when the opinion values are discrete $\pm 1, 0$. The fractions of agents having the three types of opinion values could then be written as $f_1$, $f_{-1}$ and $f_0$. It was shown analytically that at the critical point, these three fractions are equal. This is the key difference between this version and the earlier models, since in the disordered phase opinion values of opposite polarities are equally prevalent. 

 It is then straightforward to show that the order parameter behaves as
 \begin{equation}
     O=\pm \frac{\sqrt{(1-4p)}}{(1-p)}    
 \end{equation}
which implies that near $p\to p_c=1/4$, $O\sim \sqrt{p_c-p}$, giving the order parameter exponent $\beta=1/2$. This result has been confirmed with extensive numerical simulations, both for the discrete opinion values, as well as for the continuous opinion values (for which the critical point changes).

Although there is no energy function akin to a Hamiltonian in these models, from the symmetry considerations, it is seen to behave like an Ising model, at least in the mean field limit (see also Ref.~\cite{doi:10.1142/S0129183116500601}). Specifically, standard finite size scaling could be done for the order parameter, susceptibility (fluctuation of order parameter) and the Binder cumulant (see Fig. \ref{fig:contannU} for the simulation results of the version of the BChS model with continuous opinion values). For a later comparison with the exponent values of the BChS model in lower spatial dimensions, it is to be noted that the results for finite size scaling could be written with a change of variable as $N\sim L^{d}$. For the mean field, of course, this would require the knowledge of the upper critical dimension. Since the upper critical dimension of the BChS model is not known, we denote the correlation `length' exponent by $\overline{\nu}$ while writing a scaling in terms of the number of agents $N=L^d$ arrange on a $d$-dimensional lattice, implying $\overline{\nu}=d\nu$.  For example, in the case of the finite size scaling of the order parameter $O\sim L^{-\beta/\nu}F\left((p-p_c)/L^{1/\nu}\right)\sim L^{-\beta d/\overline{\nu}}F\left((p-p_c)/L^{d/\overline{\nu}}\right)\sim N^{-\beta/\overline{\nu}}F\left((p-p_c)/N^{1/\overline{\nu}}\right)$ for a $d$-dimensional lattice.

The effect of damage spreading was also studied by two different methods for the BChS model showing that the damage spreading transition takes place at $p_d$ where  $p_d <  p_c=0.25$ in the mean field case for either method~\cite{KHALEQUE2014599}.

While in the LCCC type models, only active-absorbing transitions between a dominant
state and the indifferent state can be observed,  the built-in disorder or noise in  the
interactions in the BChS type models lead to order-disorder transitions and the critical
exponents  turn  out to be the same as in the Ising model.

\subsubsection{Extreme switches and exit probability}
As mentioned before, the opinion values of either sign are possible in this model. However, as could be noted by following the dynamics of the model with discrete opinion values, if an agent is to switch their opinion value from positive to negative, or vice versa, they must first switch to the neutral opinion first.

In Ref.~\cite{PhysRevE.106.054311}, a version of the model was introduced where the magnitude of $\mu$ could be 1 or 2. In this version, only positive values of $\mu$ were considered. The interpretation for a larger value of $\mu$ would be to have a stronger influence of one agent on the other. Clearly, for $\mu=2$, the opinion value of the $i$-th agent can switch from $+1$ to $-1$ if $o_j(t)=-1$. 

If the probability for $\mu=2$ is denoted by $r$ and that for $\mu=1$ is $1-r$, then
the results are qualitatively different for $r=1$ and
$r\ne 1$. The analytical solution, which is valid in the thermodynamic limit, shows that for $r=1$ the dynamics are
quasi-conservative as the order parameter remains constant after a very short transient time. This indicates
that the system does not order fully for any initial configuration with initial order parameter less than 1. 
When a consensus is reached with either all opinions $+1$ or $-1$, one can define what is called an exit probability, which 
is a measure of the probability that the system ends up in the state towards which it was initially biased.
The
linear behavior of the exit probability is similar to what
is seen for a conservative dynamics, as for example in
the Voter model in all dimensions and the Ising Glauber
model in one dimension. This is actually quite interesting, as the present model does not strictly conserve
the order parameter; the saturation value is not exactly
equal to the initial one. But the linear behavior of the
exit probability can still occur if the saturation value of
the order parameter varies linearly with the initial value,
which was checked to be true here.
At $r=1$ as $f_0$ goes to zero very fast, it effectively renders
the system to a binary opinion model within a short time
scale with the transition rates identical to those in the
Voter model~\cite{10.1214/aop/1024404276,e15125292}. Like the voter model, here the agent
adapts the opinion of the other agent with whom she
interacts irrespective of her own opinion. It is also found
from simulations that the average consensus time is proportional
to $N$ for $r = 1$, a result valid for the mean field voter
model.

With both  $ r $ and $p \neq 0$ as parameters, the order disorder boundary in the parameter space is expressed as \cite{kathakali_new}
\begin{equation}
p=\frac{1-r}{4},
\end{equation}
while the criticality is again mean-field type.

\subsubsection{Virtual-walk in opinion space}
An interesting aspect of the dynamics was studied in Ref.~\cite{doi:10.1098/rsta.2021.0168}, where the evolution of the opinion values were associated with a virtual random walk. 
If a walker is associated to each of the individuals of the system in a virtual one dimensional  space, then the position of the $i$-th walker at time step $t+1$ in this space  can be written as
\begin{equation}
X_i(t+1)=X_i(t)+\xi_i(t+1).
\end{equation}
At each step the walker can  move to the nearest-neighbor site to its right or left  or it can remain at its present location. Then $\xi_i$ is a random number which can take  values $-1$,$0$, or $+1$. 
In this work,  the displacements $\xi$ were taken to depend on the opinion 
states. 
Two schemes were used  to implement the walk.

Scheme I 
is a Markovian process, i.e. here $\xi_i(t+1)$  depends on the present  opinion states  only: 
\begin{equation}
\xi_i(t+1)=o_i(t+1). 
\end{equation}

Scheme II
is  a non-Markovian walk where the  $\xi_i(t+1)$  depends on the present as well as the previous 
opinion states in the following way:
\begin{eqnarray}
\xi_i(t+1)& =& o_i(t+1),  ~~ {\rm 
if} ~~ o_i(t+1) = o_i(t), \nonumber \\
& = & o_i(t+1)-o_i(t), ~~ {\rm {otherwise}}. \nonumber  
\end{eqnarray}
The values of $\xi$ thus chosen are tabulated in Table~\ref{tables1}.
In either case,  $X_i(t=0) = 0$ was taken  for all $i$. 
 It is to be emphasized here that the evolution of the opinions directly involves the parameter $p$. The walks on the other hand are solely determined on the basis of the opinions in the last one or two steps and $p$ does not directly enter into the definition of the walk. 
It was found that both the walks carry the signature of the phase transition at $p_c=0.25$.
\begin{table}[h]
\caption{The table shows the values of $\xi_i(t+1)$ in the two schemes for different values of $o_i$ at times 
$t$ and $t+1$. Note that $|o_i(t) - o_i(t+1)| \leq 1$.}

\label{tables1}
\begin{center}
\begin{tabular}{ |c|c|c|c|}
\hline
 \multicolumn{2}{|c|} {} &   \multicolumn{2}{c|} {$\xi_i(t+1)$}\\ 			

\cline{1-4}

 $o_i(t)$  & $o_i(t+1)$ &			 Scheme I 	& 	Scheme II	 	 \\

\hline
 1			&	1  		&	 1	& 	1		\\
\hline
 1			&	0  		&	 0	& 	-1		\\
\hline
0 & 0 & 0 & 0\\
\hline
0 & 1 & 1 & 1\\
\hline
0 & -1 & -1 & -1\\
\hline
-1 & 0 & 0 & 1\\
\hline
-1 & -1 & -1 & -1\\
\hline

\end{tabular}
\end{center}

\end{table}

\subsubsection{Public-private opinions}
It was noted that the publicly expressed opinion might differ from the privately held beliefs (see e.g., Ref.~\cite{KING198190}). Due to peer pressure and/or political purposes, an agent can have a private and a public opinion value, which might differ in magnitude as well as in sign. In Ref.~\cite{doi:10.1098/rsta.2021.0169}, a version of the kinetic exchange model for opinion was considered, where the two types of opinion values were treated separately. Particularly, the public opinion values were allowed to follow the exchange rule mentioned before, while the private opinion values, denoted by $P_i(t)$, followed
\begin{equation}
    P_i(t+1)=P_i(t)+k(o_i(t+1)-o_i(t))
\end{equation}
where $k$ is a parameter. This dynamics are not reflected in the public opinion, until the public opinion value differs from the private opinion value of a particular agent by more than a tolerance parameter $\delta$, such that
\begin{equation}
    o_i(t)=\mathrm{sgn}(P_i(t))
\end{equation}
if $|o_i(t)-P_i(t)|>\delta_i$. 
One could then measure two order parameters $O(t)$ for the public opinion values as before and $Q(t)=\frac{1}{N}|\sum\limits_i P_i(t)|$ for the private opinions. It was shown numerically that a transition to disorder happens for a value of $p$, depending on the values of $k$ and $\delta$, but a statistically significant difference between the two kinds of opinion values persists in all phases of the dynamics. While the critical behavior is Ising-like for high values of $\delta$, for low values of $\delta$, it was seen to follow non-Ising exponent values. This is a signature of the non-Ising nature of the model, which we shall come back to in the later sections.

\subsubsection{Contrarians and zealots}
In a society, not all agents would follow an opinion `exchange' as written in Eq. (\ref{gen_eq}). Indeed, they might not enter into an `exchange' at all, i.e., they can retain their opinion values indefinitely. Alternatively, they could behave contrary to the norm, i.e. take an opinion value opposite to that dictated by the interaction rule. 

A group of agents not following the `rules' defined for most people in the society have been considered before in many variants of opinion formation models (see e.g., Ref.~\cite{GALAM2004453}). Here, we revisit the studies that looked into the effect of such group(s) of agents in the BChS model.

In Ref.~\cite{PhysRevE.89.013310} a parameter denoting the fraction of inflexible agents was introduced to study how it affects the opinion formation (see also Refs.~\cite{PhysRevE.92.062122,GAMBARO2017465}). Such a fraction of agents do not change their opinions in any type of interaction. Introduction of this fraction lowers the value of $p_c$. The inflexible agents could either be chosen randomly, or could only belong to  either of the extreme opinions, or both. The resulting phase boundary depends on this choice, but the universality remains unaffected. In contrast to the BChS and the LCCC model in Ref.~\cite{PhysRevE.86.061127} the conviction  parameter $\lambda$ was chosen as a random variable with discrete values (either $0$, $1$ or $-1$), which gives rise to a two parameter model. Such a modification does not lead to any change in the universality class. However, the phase boundary shows that with the presence of $\lambda =0 $ or $-1$ would lead to a lower $p_c$ value.  A similar  model was proposed in Ref.–\cite{Vieira_2016}. Here, additionally, the provision of independent selection of opinion by the agents was considered, irrespective of the states of their own and the agent they are interacting with.  The critical behavior was found to be similar to the mean field Ising model. There are also similar kinetic exchange opinion dynamics models~\cite{CROKIDAKIS20141683,VIEIRA20162632} which eventually produce order-disorder transitions with mean field Ising critical exponents. In Ref.~\cite{Crokidakis_2013} the relaxation behavior of a three-state ($\pm 1, 0$) opinion dynamics model on a square lattice was studied. The evolution of the states of the agents is governed by the dynamical rules similar to the voter model~\cite{RevModPhys.81.591}. In addition to this, Ref.~\cite{Crokidakis_2013} considered a noise in the system which can change the opinion of any agent to the neutral state. A similar model with a community structure was considered in Ref.~\cite{PhysRevE.100.032312}. In this study, the value of  $\mu =1$ if the interacting agents belong to the same community and $-1$ otherwise. The study accounted for several parameters relevant to the community structure and links, and eventually identified the ordered and disordered phases. The role of the  inflexible agents for $p=0$ was also studied in this work. In refs.~\cite{PhysRevE.95.042308,doi:10.1142/S2424942417400011} a parameter $T$ was introduced, which effectively plays a role of ``social temperature", and captures the degree of randomness in the behavior of agents. The dynamical equation of the BChS model was altered by multiplying factor $1/T$ in the RHS of Eq.(\ref{bcs_dyn}) and finally a hyperbolic tangent was taken on it. The effect of this ``social temperature" manifests in the existence of three phases at $p = p_c$; symmetric (opinions are symmetrically distributed between $+1$ and $-1$), asymmetric (opinions are asymmetrically distributed between $+1$ and $-1$) and neutral (an absorbing state, the distribution is peaked about zero) in the $p_c-T$ plane. Interestingly, $p_c$ shows a slow rise with temperature for low temperatures, however, as the temperature is increased beyond a certain value, both the symmetric and asymmetric phases transit to the neutral one. Ref.~\cite{PhysRevE.95.042308} reported that the critical behavior of the absorbing phase transitions belongs to directed percolation universality class.

\subsection{Effect of topology: Lattices and Networks:}
Under most circumstances, there are no realistic restriction in the interaction or opinion exchange due to spatial constraints usually seen for physical models. However, there could be other types of constraints that could eventually give rise to a restricted neighborhood of interaction for an agent. For example, it is widely known that social networks often have scale-free degree distribution~\cite{PhysRevE.66.035103}.  Similarly, it is also known that there is a natural bound in the human brain for maintaining friendship~\cite{DUNBAR1992469}, which means a fully connected graph may not be the ideal topology to implement an opinion dynamics model. Of course, the fully connected graph is where the mean field approximation is exact, which is analytically tractable. But naturally the questions of dimensional dependence or more generally the topological dependence of the critical exponents could be raised. 

In view of this, the BChS model and its variants have been studied in lower dimensions (regular lattices), quasi-periodic lattices and on various networks, where the edges represent interaction possibilities and the nodes are the locations of the agents. We have discussed some of these instances in the earlier subsection. However, here we focus on the systematic studies concerning the critical exponent values and their variations due to topology.

\subsubsection{BChS model on lattices: Regular and quasi-periodic}
The study of the BChS models and for that matter any other opinion dynamics model on a regular lattice is primarily motivated by the assertion of its universality class and thereby determining the lower and upper critical dimensions. It could also have implications in growth, dynamics and coarse-graining of similarly opinionated neighborhoods~\cite{PhysRevE.96.032303} that do form for various different reasons. 

In Ref.~\cite{PhysRevE.94.062317}, the BChS model was studied in two and three-dimensional lattices numerically (there is no transition at finite noise in one dimension). Both the cases of $\mu_{ij}(t)$ having discrete and continuous values were considered. Correspondingly, the opinion values are discrete $\pm 1, 0$ or continuous.  While the critical points depend on this, the critical exponent values do not depend on discrete or continuous values of $\mu_{ij}(t)$ or $o_i(t)$.

In Ref.~\cite{doi:10.1142/S0129183120500126}, the BChS model was studied on quasi-periodic lattices (see also Refs.–\cite{10.3389/fphy.2017.00047,e19090459}). The authors considered Penrose, and Ammann-Beenker lattices. They also considered 7-fold and 9-fold quasi-periodic lattices. In general, it is expected that the universality class is not altered in quasi-periodic lattices. Here, the authors also confirm the same, i.e, the exponent values remain the same as the two-dimensional lattice. 

In Ref.~\cite{doi:10.1142/S0129183117501236}, the model was studied on triangular, kagome and honeycomb lattices. It is interesting to note that in this case, the exponent values were slightly different from those seen for two-dimensional regular lattices.  In Table~\ref{tab:exp}, the exponents values are summarized.

\begin{table}[h]
\caption{Comparing the critical exponents of the model studied, with Ising model in different 
dimensions. Mean field exponents for the model are taken from Ref.~\cite{BISWAS20123257}, 
while exponents of Ising model are taken from Ref.~\cite{stanley1971introduction}  ($d=2$, 
exact results) and Ref.~\cite{goldenfeld} ($d=3$).}
\centering
\resizebox{\textwidth}{!}{ 
\begin{tabular}{|l|c|c|c|c|c|c|}
\hline
dimension $d$ & lattice & $\mu$ & $p_c$ & $\nu$ &  $\beta$ & $\gamma$ \\ \hline

mean field & & discrete  & $\frac{1}{4}$ (exact); $0.250\pm 0.001$~\cite{BISWAS20123257}  
& $\bar{\nu} =2.00\pm0.01$ & $\frac{1}{2}$ (exact); $0.50 \pm 0.01$~\cite{BISWAS20123257} & 
$1.00\pm0.05$~\cite{BISWAS20123257}  \\ \hline

mean field & & continuous  & $0.3404\pm0.0002$~\cite{BISWAS20123257} & 
$\bar{\nu}=2.00\pm0.01$~\cite{BISWAS20123257}& 
$0.50 \pm 0.01$~\cite{BISWAS20123257}  &  $1.00\pm0.05$~\cite{BISWAS20123257}\\ \hline

$d=2$ & square & discrete  & $0.1340\pm0.0001 $~\cite{PhysRevE.94.062317} & $0.99\pm0.01$~\cite{PhysRevE.94.062317} & $0.122 \pm 0.002$~\cite{PhysRevE.94.062317} & 
$1.75\pm 0.01$~\cite{PhysRevE.94.062317}  \\ \hline 

$d=2$ & square & continuous & $0.2266\pm0.0001$~\cite{PhysRevE.94.062317} & $0.99\pm0.01$~\cite{PhysRevE.94.062317} & 
 $0.125 \pm 0.001$~\cite{PhysRevE.94.062317} & $1.75\pm0.01$~\cite{PhysRevE.94.062317}  \\ \hline 

$d=2$ & triangular & continuous  & $0.123 \pm 0.0006$ \cite{doi:10.1142/S0129183117501236} & $0.97 \pm 0.0008$\cite{doi:10.1142/S0129183117501236} & 
 $0.14 \pm 0.005$\cite{doi:10.1142/S0129183117501236} & $1.58\pm0.002$\cite{doi:10.1142/S0129183117501236}  \\ \hline

 $d=2$ & honeycomb & continuous & $0.115 \pm 0.0004$ \cite{doi:10.1142/S0129183117501236} & $1.14 \pm 0.009$\cite{doi:10.1142/S0129183117501236} & 
 $0.19 \pm 0.002$\cite{doi:10.1142/S0129183117501236} & $1.81\pm0.006$\cite{doi:10.1142/S0129183117501236}  \\ \hline

 $d=2$ & kagome & continuous  & $0.068 \pm 0.0003 $\cite{doi:10.1142/S0129183117501236} & $1.16 \pm 0.002$\cite{doi:10.1142/S0129183117501236} & 
 $0.16 \pm 0.002$\cite{doi:10.1142/S0129183117501236} & $1.89\pm0.005$\cite{doi:10.1142/S0129183117501236}  \\ \hline

$d=2$ & Penrose quasiperiodic & continuous  & $0.2293 \pm 0.00005 $\cite{doi:10.1142/S0129183120500126} & $1$\cite{doi:10.1142/S0129183120500126} & 
 $1/8$\cite{doi:10.1142/S0129183120500126} & $1$\cite{doi:10.1142/S0129183120500126}  \\ \hline

$d=2$ & Ammann-Beenker quasiperiodic &  continuous & $0.2299 \pm 0.00005$\cite{doi:10.1142/S0129183120500126} & $1$\cite{doi:10.1142/S0129183120500126} & 
 $1/8$\cite{doi:10.1142/S0129183120500126} & $1$\cite{doi:10.1142/S0129183120500126}  \\ \hline

$d=2$ & Seven-fold quasiperiodic & continuous  & $0.2290 \pm 0.00005$\cite{doi:10.1142/S0129183120500126} & $1$\cite{doi:10.1142/S0129183120500126} & 
 $1/8$\cite{doi:10.1142/S0129183120500126} & $1$\cite{doi:10.1142/S0129183120500126}  \\ \hline

$d=2$ & Nine-fold quasiperiodic & continuous  & $0.2290 \pm 0.00005$\cite{doi:10.1142/S0129183120500126} & $1$\cite{doi:10.1142/S0129183120500126} & 
 $1/8$\cite{doi:10.1142/S0129183120500126} & $1$\cite{doi:10.1142/S0129183120500126}  \\ \hline

$d=3$ & cubic & discrete  & $0.1992 \pm 0.0002$ & $0.63\pm0.01$  & $0.310 \pm 0.002$ & 
$1.255\pm0.005$ \\ \hline

$d=3$ & cubic & continuous & $0.2854\pm0.0001$ & $0.63\pm0.01$ & $0.310 \pm 0.002$ &
$1.26\pm0.01$ \\ \hline \hline 

mean field Ising & & & & $\nu=\frac{1}{2}$; $d=4$  (exact)~\cite{stanley1971introduction}  & 
$\frac{1}{2}$  (exact)~\cite{stanley1971introduction} & $1$  
(exact)~\cite{stanley1971introduction}\\ \hline 

$d=2$ Ising & & & & $1$  (exact)~\cite{stanley1971introduction}  & 
$\frac{1}{8}$  (exact)~\cite{stanley1971introduction} & 
$\frac{7}{4}$ (exact)~\cite{stanley1971introduction}\\ 
\hline 

$d=3$ Ising  & & & & $0.630\pm 0.002$~\cite{goldenfeld} & 
$0.3250\pm 0.0015$~\cite{goldenfeld} & $1.2405\pm 0.0015$~\cite{goldenfeld}\\ 
\hline

\end{tabular}
}
\label{tab:exp}
\end{table}

\subsubsection{Networks}
As mentioned before, an important aspect of the study of opinion dynamics model is its implementation on realistic social network structures viz., scale free networks. The BChS model was studied on different network topologies. Particularly, in ref.~\cite{e25020183}, the critical behavior of the BChS model simulated on a directed Barab\'{a}si-Albert Networks (DBAN) (see also, Refs.~\cite{e21100942,LIMA2021125834,Alves_2020}) was investigated. It was shown that the value of $p_c$ as well as the ratios of $\beta/\nu$ and $\gamma/\nu$ change non-monotonically with the connectivity. It was also reported that the universality class of the BChS model on DBAN is same as of majority-vote model (MVM). In ref.~\cite{RAQUEL2022127825}  the nonequilibrium BChS model on Erd\"{o}s-R\'{e}nyi random network (ERRG) and directed ERRG random network were studied. Their numerical results indicate that the critical behavior of the BChS model on such graphs is different from the MVM realized on same networks. The universality class is also different from the equilibrium Ising universality class. 

\section{Applications}
\subsection{Coarse grained information: US elections}
\begin{figure}
\includegraphics[width=14cm]{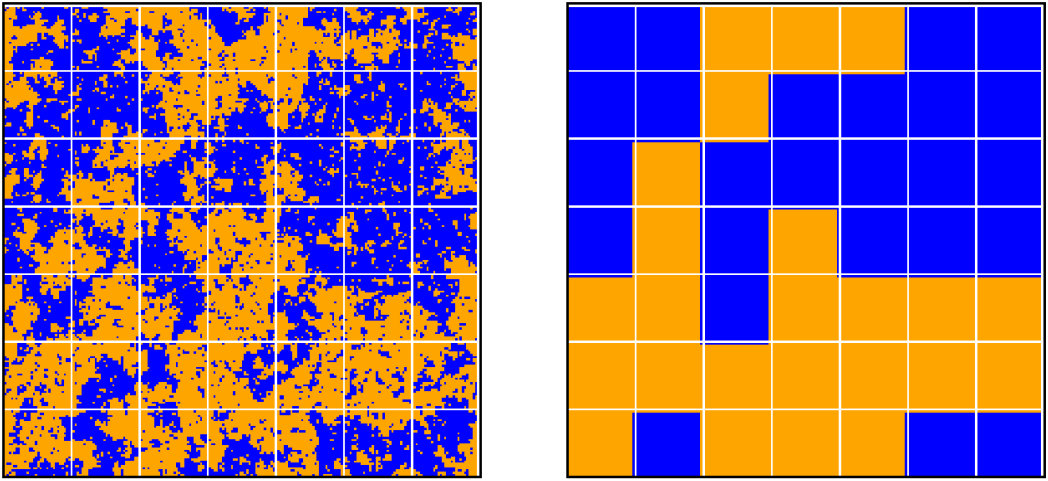}
\caption{A schematic representation of coarse graining in the Ising model. The left hand side figure shows the spatial configuration of the up (orange) and down (blue) values on a square lattice and the coarse graining boundaries are indicated. The right hand side figure shows a coarse grained picture. Although initially there were more up (22382) than down (21718) values, after the coarse graining there are more down (25) than up (24) blocks. This is an instance where coarse graining flips the sign of the average order parameter. This particular case uses two dimensional Ising model. But a similar picture could arise form the BChS and similar other models. Taken from~\cite{PhysRevE.96.032303}.}
\label{snapshot}
\end{figure}

The 2016 US Presidential election revealed an intriguing aspect: the candidate who received a greater share of the popular vote lost the election. This can be attributed to the electoral college system of the US, where in most states the winner of the popular vote in a state wins all of its electoral college votes as well. Essentially, this is a process of coarse graining. While the renormalization group theory of critical points shows that coarse graining near a critical point does not change the scaling behavior of a system, the sign of the order parameter can be flipped due to the coarse graining. In the case of the 2016 US elections, if we consider the electoral college as a spatial coarse graining of the popular vote, a flip of the sign of the order parameter (average opinion value) occurred (see Fig. \ref{snapshot}). The probability of such an event (flip) happening is significant when the underlying system (here the popular vote) is near a critical point i.e., having no clear winning opinion and strong spatial correlation in the spatial organization of the opinion states~\cite{PhysRevE.96.032303}. Indeed, there are four instances of the minority candidate winning in the US presidential elections: 1876, 1888, 2000 and 2016.
The coarse graining process was applied to the kinetic exchange opinion model studied on a square lattice, which involved examining the time series of order parameter values before and after the process. During this process, the behavior resembled that of a noisy channel, where certain values may have been flipped, resulting in a change of sign. One can subsequently quantify the loss of information from a measurement of  the mutual information between the two time series~\cite{PhysRevE.96.032303}.

\begin{figure}
\includegraphics[width=10.0cm]{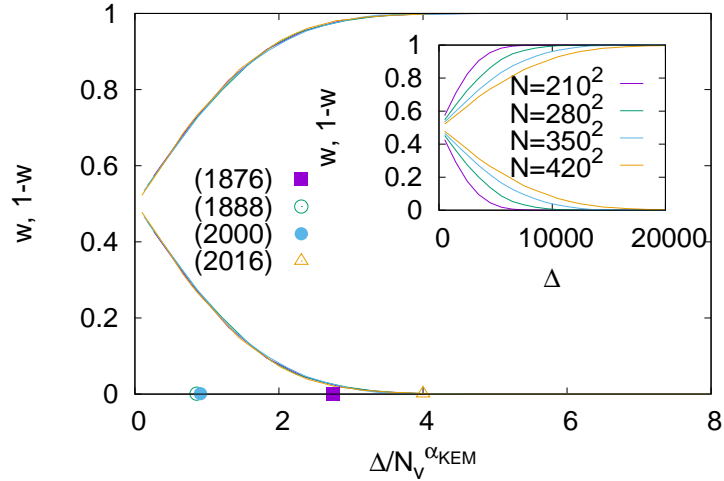}
\caption{The finite size scaling of the probability of the minority candidate winning $w$ as a function of the difference in the population of the two non-zero opinion values $\Delta$. The inset shows the unscaled data. The x-axis values of the four cases of the minority candidate winning in the US presidential elections are indicated using the value of $\Delta$ from historic voting data and the value of the exponent same as that obtained from the above mentioned finite size scaling. The location of the x-axis values indicate that there has always been a significant chance of the minority candidate winning in these elections. Taken from Ref.~\cite{PhysRevE.96.032303}.}
\label{minwin_2dbcs_p0.12}
\end{figure}
Particularly, if $\Delta$ is the difference between the two signs of the (extreme) opinion values in the BChS model in two dimensions, then from the time series of the order parameter and that of the coarse grained lattice (with 49 blocks) one can estimate the fraction of the times when a flip of sign have occurred due to coarse graining. This is the probability of the minority candidate winning $w$. When measured near the critical point $p_c\approx 0.12$ of the two dimensional BChS model, this quantity shows a finite size scaling (see Fig. \ref{minwin_2dbcs_p0.12}) of the form $w=G\left(\Delta/N_v^{\alpha_{BChS}}\right)$, where $N_v$ is the number of agents having non-zero opinion values (about 80\% of the total population for the BChS model in two dimensions near the critical point) and $\alpha_{BChS}\approx 0.7$~\cite{PhysRevE.96.032303}.

As noted above, the process of coarse graining essentially implies a loss of information, much like a noisy channel~\cite{shannon1948mathematical}.  Here the input signal is the sign of the majority of $N$ agents (denoted by $\mathcal{N}$) and the output signal is the sign of the majority of the coarse grained system ($\mathcal{M}$). The mutual information ($I$) transferred from the input to the output is then given by
\begin{equation}
 I(\mathcal{N},\mathcal{M})=H(\mathcal{N})+H(\mathcal{M})-H(\mathcal{N},\mathcal{M}),
\end{equation}
where
\begin{eqnarray}
H(X) &=& -\sum\limits_{i\in\{0,1\}}p(X=i)\log p(X=i),\nonumber \\
H(X,Y) &=& -\sum\limits_{i,j\in\{0,1\}} p(X=i \wedge Y=j)\log p(X=i \wedge Y=j) \nonumber
\end{eqnarray}
where $p(X=i)$ is the probability of input being $i$ and so on.  
The relative mutual information $R$ is then given by~\cite{szcz, agni}
\begin{eqnarray}
R(\mathcal{N},\mathcal{M})&=&\frac{H(\mathcal{N})+H(\mathcal{M})-H(\mathcal{N},\mathcal{M})}{[H(\mathcal{N})+H(\mathcal{M})]/2} 
\label{eq:rmi}
\end{eqnarray}
which is a measure of the reduction in the uncertainty of the input, given the knowledge about
the realization of the output, relative to the average uncertainty of the input and output.
The value of $R$ is 1 below the critical point, which implies that the output is fully predictable from the input. Above the critical point, $R$ sharply drops to zero, where all information is lost.

\subsubsection{Block size dependence}
\begin{figure}
\includegraphics[width=10.0cm]{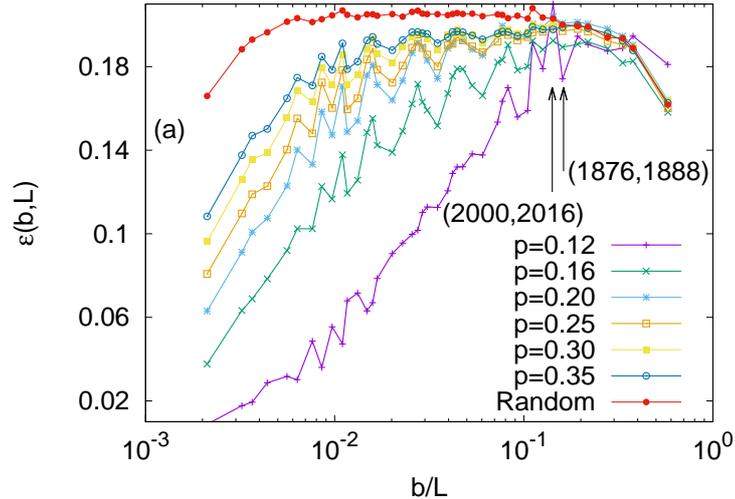}
\caption{The probability of the minority candidate winning is plotted against the coarse graining block size $b$ scaled by the linear system size, in the BChS model for different values of $p$. The vertical arrows indicate the positions of the the x-axis where the actual minority candidate winning incidences have occurred, if the $b/L$ ratio is taken as the electoral college size and voter size ratio in the US for those four years. It seems that almost for any value of $p$, the electroral college sizes are such that the minority candidate winning probability is significantly high. Taken from Ref.~\cite{BISWAS2021125639}.}
\label{min_cand}
\end{figure}

Furthermore, this loss of information is found to be dependent on the size of the coarse graining blocks~\cite{BISWAS2021125639}. 
It can be easily understood that the limit of unit block size and a system wide block size would give back the original system, implying no loss in information. However, for intermediate sizes, there will be loss of information which will be maximum for a particular size. Interestingly, at the current state of the block (states of USA) sizes, the loss is near the maximum (see Fig. \ref{min_cand}). This may call for proper attention, not  the least while making pre-poll predictions of such results. 

\subsection{Brexit: A long route to consensus}
The question of the UK leaving the European Union has been a debated topic for half a century. The reason that the issue remained active in the UK politics (or for that matter the EU politics) is the lack of consensus regarding the two choices. Of course this is an interesting issue which was addressed in opinion dynamic models elsewhere (see e.g., \cite{brx1,brx2}), but in the present context, it can be thought of as a binary choice opinion evolution.

In the BChS model, if the noise parameter is set to zero ($p=0$), then coarsening will happen (similar to $T=0$ in the Ising model) and the system will eventually go to a consensus (all up or all down) state, at least on the lower spatial dimensions. If the initial state is disordered, then there will be competition between up and down domains. Interestingly, the domain boundaries will be separated by neutral agents. This is in contrast with what one observes in Ising model. 

\begin{figure}[tbh]
\center
\includegraphics[width=14cm, keepaspectratio]{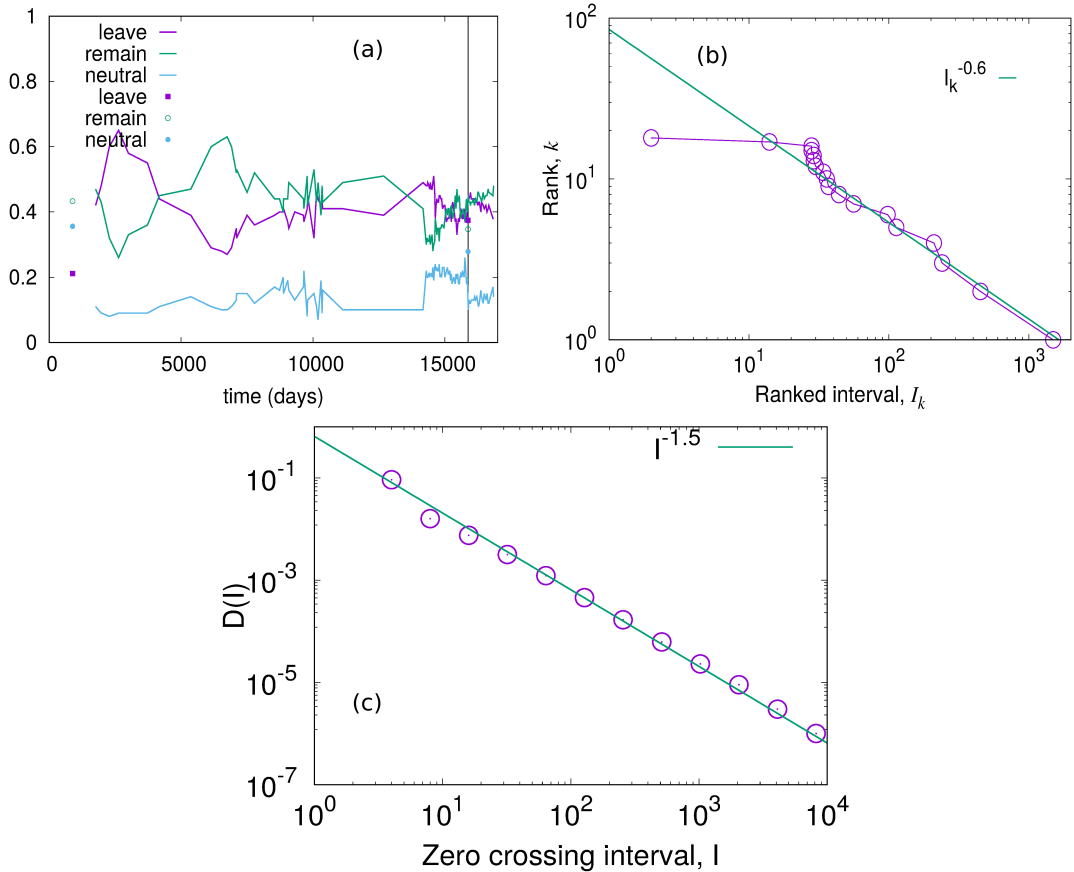} 
\caption{Comparisons between Brexit and the BChS model data.  (a) The figure shows various opinion surveys and referendum on the question of the UK leaving the EU from the date of its joining (1 January, 1973 to the then European Communities) as the origin ($t=0$). The
vertical line denotes the time of the last referendum (23 June, 2016). (b) The rank-plot of the interval of the zero-crossing of the net opinion value (difference
between remain and leave fractions) is shown. The tail of the rank plot shows an exponent close to $-0.60\pm0.02$. (c) The distribution $D(I)$ of intervals of zero crossing for the BChS model (circles) and its fitting with an exponent $-1.5$, implying that the cumulative (seen for the real data in (b)) would give an exponent value $-0.5$. Taken from Ref.~\cite{PhysRevE.102.012316}.}
\label{brexit_data}
\end{figure}

It is noted~\cite{PhysRevE.102.012316} that approximately one-third of the configurations go to a trapping state, where the dynamics are not frozen, but the domain sizes of opposite signs remain comparable for a very long time. These configurations take a longer time (different scaling with system size) than the remaining two-third, which reach consensus much more quickly. If averaged over all configurations, this would be reflected as a two-stage consensus process, similar to what was also seen in the voter model. Such a longer route to the consensus has analogs in society, where some contentious issues divide people in such a way that finding an overall consensus may remain elusive for decades. The question of the UK leaving the European Union is/has been one such issue. Interestingly, there are data for opinion polls going back many decades. These data show that the overall population remained divided almost equally on this issue, with remain/leave campaigns marginally gaining over each other without any clear dominance. Indeed, the distribution of the zero crossing could be plotted and compared with the same in the BChS model in two dimensions with $p=0$ (see Fig. \ref{brexit_data}). The scaling behavior of the theoretical results and the real data show promising agreement.

\subsection{Tax evasion dynamics}
An interesting application of the three state kinetic opinion formation is in the case of tax evasion dynamics~\cite{tax_lima,CROKIDAKIS2014321}. There have been earlier studies on tax evasion dynamics with opinion models, particularly the Zaklan model~\cite{zaklan}, where two opinion states were considered, representing the tax payers and the tax evaders. A similar parallel is drawn for the BChS model as well, i.e. the opinion value $o_i(t)=+1$ would imply that the $i$-th agent is a tax payer at time $t$, and $o_j(t)=-1$ would imply that the $j$-th agent is a tax-evader at time $t$.  However, it is interesting to note the effect of the neutral agents with opinion values 0, who represent the undecided fraction of agents. They can change their state to tax payers or tax evaders depending on their subsequent interactions. 

A punishment rule is then applied, which means that a randomly selected fraction of the tax evaders are audited and changed to the tax payers state for some subsequent time steps. After that time period, they can again participate in the opinion dynamics as before and can switch to tax evaders state. 

In the ordered phase of the model, the punishment rule does not affect the state of the system significantly. However, in the disordered state, where all three fractions are usually present in the same fraction, the enforcement of the punishment rule can significantly reduce the tax evader fraction.

\section{Discussion and Conclusion}

The kinetic exchange models of opinion formation have made significant  contribution in
understanding how a society reaches or does not reach a collective decision. 
In this review, we have provided an up-to-date overview of the opinion formation models within this category, as the research in this area remains quite active.
 Our focus has been mainly on the Biswas-Chatterjee-Sen (BChS) model~\cite{BISWAS20123257} proposed in 2012 and its variants and later developments.

An interesting issue in this class of opinion formation models is the existence of phase transitions between symmetric and symmetry broken phases governed by suitable driving parameters. In the BChS model, where a negative interaction can occur between agents, such a phase transition was shown to take place above a critical fraction of negative interactions on a fully connected network. Later more parameters have been introduced to include various features like the  presence of  inflexible or contrarian
agents, independent opinion formation, random opinion changes, extreme switches, etc., all of which occur with certain probabilities.  These  usually produce additional noise in the system. We have discussed these cases in sec IIA.

An important aspect in studies on social phenomena is the topology of the network on which the agents are placed; this decides the  connectivity of the agents.  The results are strongly dependent
on the topology. While exact results are only available for the mean field cases for the BChS model,  in later works, approximate results and simulations have been
made  on finite dimensional regular networks, random graphs, scale free
networks etc. A short review of such models, some of which  also incorporated other possible sources
of noise, has been made in sec IIB.

The nature of the phase transition and  the universality class (when one has a continuous phase transition) have also been investigated for these models.
The LCCC type models may be regarded as predecessors of the BChS model. In the former, the disordered phase is absorbing. But the noise parameter in the BChS type models makes it close to the Ising universality class.
Therefore, in most of the cases, one finds the critical exponent values to
be very close to those of the Ising model,  although whether in general
the BChS model belongs to the Ising critical class has not yet been  established.
This is because, for example, in the mean field case, one has to assume an effective dimension equal to 4 in order to get correspondence with respect to all the critical
exponents.

The kinetic exchange model including the BChS model  differs intrinsically from the Ising model and binary opinion dynamics models like the Voter model as it allows more than two opinion states even in the discrete version. As a result there can be  an absorbing state also
when all opinions become zero - it is disordered as the order parameter is zero, on the other
hand it differs strongly from the symmetric phase.
In the cases where one has transition to this  absorbing state,   a directed percolation (DP) like universality has also been claimed. These discussions have also been included in sec IIB.

However, for an opinion dynamics model  to be truly useful and acceptable, one needs to show that it works
reasonably well when compared with real data. The success of the BChS  model lies in the fact that one can indeed get consistency with real data in at least two cases, namely the  US Presidential election and Brexit, using
appropriate topology and parameters. These applications have been discussed in detail in sec III, in the context of coarse graining in the US presidential election and the subsequent probability of a minority candidate winning (IIIA),  the scaling behavior of the consensus time in binary choices applied to the case of Brexit (IIIB) and application to tax evasion models (IIIC).

In short, we have consolidated here the results available for  the BChS model and its
modified versions as of now,  and expect to see research in several directions based on these models in future.

\section*{Acknowledgement}
The authors acknowledge collaborations at various stages with Kathakali Biswas,  Anindya S. Chakrabarti, Anirban Chakraborti,  Anjan Kumar Chandra, Krishanu Roy Chowdhury,  Asim Ghosh, Abdul Khaleque, Mehdi Lallouache, Subhadeep Roy and  Surajit Saha.
%

\end{document}